# Observations of the Quadrantid meteor shower from 2008 to 2012: orbits and emission spectra

**José M. Madiedo[1, 2], Francisco Espartero[2,3], Josep M. Trigo-Rodríguez[4], Alberto J. Castro-Tirado[5], Pep Pujols[6], Sensi Pastor[7], José A. de los Reyes[7] and Diego Rodríguez[8]**

[1] Departamento de Física Atómica, Molecular y Nuclear. Facultad de Física. Universidad de Sevilla. 41012 Sevilla, Spain.

[2] Facultad de Ciencias Experimentales, Universidad de Huelva, 21071 Huelva, Spain.

[3] Observatorio Astronómico de Andalucía, 23688 La Pedriza, Alcalá la Real, Jaén, Spain.

[4] Institut de Ciències de l'Espai (CSIC-IEEC), Campus UAB, Facultat de Ciències, Torre C5-parell-2ª, 08193 Bellaterra, Barcelona, Spain.

[5] Instituto de Astrofísica de Andalucía, CSIC, Apt. 3004, 18080 Granada, Spain.

[6] Grup d'Estudis Astronòmics (GEA) and Agrupació Astronòmica d'Osona, Barcelona, Spain.

[7] Observatorio Astronómico de La Murta. Molina de Segura, 30500 Murcia, Spain.

[8] Guadarrama Observatory (MPC458), Madrid, Spain.

**ABSTRACT**

The activity of the Quadrantids in January during several years (2008, 2010, 2011 and 2012) has been investigated in the framework of the SPanish Meteor Network (SPMN). For this purpose, an array of high-sensitivity CCD video devices and CCD all-sky cameras have been used to obtain multi-station observations of these meteors. These allowed us to obtain precise radiant and orbital information about this meteoroid stream. This paper presents a large set of orbital data (namely, 85 orbits) of Quadrantid meteoroids. Most meteors produced by these particles were recorded during the activity peak of this shower. Besides, we discuss four Quadrantid emission spectra. The tensile strength of Quadrantid meteoroids has been also obtained.

## 1. INTRODUCTION

The Quadrantid meteor shower was first identified in 1835 (Fischer 1930, Lovell 1954), and no records of this shower earlier than the beginning of the 19th century appear to exist (Williams and Colander-Brown 1998). Its activity extends from about December 31 to January 6, peaking around January 4 (Jenniskens 2006). Besides, this peak is very short, being less than a day (Shelton 1965, Williams et al. 1979).

For many years the parent of the Quadrantid stream was unknown though many suggestions were made (see Williams & Collander-Brown 1998 for a list). The two most likely candidates were Comet 96P/Machholz (McIntosh 1990) and Comet 1490I (Hasegawa 1979, Williams & Wu 1993). Jenniskens (2004) showed that asteroid 2003EH1 and the Quadrantids had an exceedingly similar orbit and that there must be a generic relationship between them. In that paper Jenniskens suggested that the fragmentation of Comet 1490I at some later date might have led to the formation of both 2003EH1 and the core of the Quadrantids. This suggestion was investigated further by Williams et al. (2004). Neslušan et al. (2013) have argued that the Quadrantids and 2003EH1 resulted from a break-up of Comet 96P/Machholz, thus explaining both the core and the broad component of the Quadrantids. Jopek & Williams (1993) took the interrelationship idea further, suggesting that a proto-Machholz fragmented to form both the present day Machholz and 1490I, with a further fragmentation producing 2003 EH1 and the Quadrantids.

The observations of the Quadrantids are not abundant. Thus, despite having the highest zenithal hourly rate (ZHR) of all annual meteor showers (about 130 meteors per hour), its short activity period and frequent unfavourable weather conditions in early January pose important difficulties to the observation of this shower. This short activity period makes the total amount of observing time very short each year. Hence the advantage of accumulating several years of observations. Wu and Williams (1992) published the first large set of precisely reduced orbits of Quadrantid meteoroids, which consisted of 118 orbits. Jenniskens et al. (1997) produced a list of 64 Quadrantid orbits. A continuous monitoring of meteor and fireball activity from sites offering good probabilities of optimal weather conditions is very convenient in order to analyze this stream. One of these locations within continental Europe is the Iberian Peninsula, particularly the South of Spain. This has provided the opportunity to observe the Quadrantids in recent years from several meteor stations operating in the framework of the SPMN. Here we provide 85 orbits obtained for multi-station Quadrantid meteors imaged between 2008 and 2012. Four emission spectra produced during the ablation of particles belonging to this meteoroid stream are also presented and discussed.

## 2. INSTRUMENTATION AND METHODS

Several meteor observing stations located in Spain were involved in the monitoring of the activity of the Quadrantid meteor shower from 2008 to 2012. Their locations are listed in Table 1. Except for station #9, which operates an all-sky slow-scan CCD camera, these stations employ an array of low-light CCD video cameras (models 902H and 902H Ultimate, from Watec Co.) to obtain meteor atmospheric trajectories and meteoroid orbits (Madiedo & Trigo-Rodríguez 2008; Trigo-Rodríguez et al. 2007). The video cameras are equipped with a 1/2" Sony interline transfer CCD image sensor with their minimum lux rating ranging from 0.01 to 0.0001 lux at f1.4. Aspherical lenses are employed. Their focal length ranges from 6 to 25 mm and the field of view covered by each device ranges from 62 x 50 to 14 x 11 degrees. In this way, different areas of the sky can be monitored by every camera and point-like star images are obtained across the entire field of view. These cameras generate interlaced imagery according to the PAL video standard, at a rate of 25 frames per second and with a resolution of 720x576 pixels. Most of our video stations develop a continuous monitoring of the night sky and work in an autonomous way by means of the MetControl software (Madiedo et al. 2010, Madiedo 2014). However, station #2 (Cerro Negro) is a mobile system which is set-up when necessary in a dark countryside environment at about 60 km north from Seville. A detailed description of the all-sky CCD system operating at station #9 can be found in Trigo-Rodríguez et al. 2007. Aspherical fast lenses with focal lengths ranging from 4 to 12 mm and focal ratios between 1.2 and 0.8 were used. In this way, different areas of the sky were covered by every camera and point-like star images were obtained across the entire field of view. With this configuration we can image meteors with an apparent magnitude of about 3 ± 1. The images taken by the cameras are sent to an array of PC computers which are automatically synchronized by means of GPS devices. In this way, meteor recording times are known with an accuracy of 0.1 seconds. A more detailed description of the operation of these systems can be found in (Madiedo & Trigo-Rodríguez 2008; Madiedo et al. 2010). To reduce the images containing meteor trails we have followed the procedure described in (Madiedo et al. 2013a). Thus, at the end of every observing session data recorded during the night were automatically compressed and sent to our FTP server. This is not the case for the mobile station at Cerro Negro, where the video files were manually saved to the server's hard disk. Once the images recorded by every station were stored on the FTP server, another software package identified trails that were simultaneously recorded from at least two different stations. A copy of these multi-station data was placed in a separate folder where the video frames on each video file were co-added in order to increase the number of stars available for the astrometric measurement. A composite image showing the whole meteor trail was also generated for each event. Then, an astrometric measurement was done by hand in order to obtain the plate (x, y) coordinates of the meteor along its apparent atmospheric path from each station. These astrometric measurements were introduced in the AMALTHEA software (Trigo-Rodríguez et al. 2009a,b; Madiedo et al. 2011), which was developed by the first author and employs the methods described in (Ceplecha 1987) to obtain the atmospheric trajectory of the meteor and the orbital elements of the progenitor meteoroid.

In order to count meteors occurring during our survey in 2011, a forward-scatter radio system operating at a frequency of 143.05 MHz was operated. This was located in Collado Villalba (Madrid), and employed a 2.15 dBi discone antenna (model Diamond D-130) connected to a Yaesu VR 5000 radio receiver. This device received the reflections from the Grand Réseau Adapté à la Veille Spatialle (GRAVES) radar, located in Dijon, France (http://www.onera.fr/dcps/graves).

Another aim of this research was to obtain emission spectra of Quadrantid meteors. To accomplish this task, holographic diffraction gratings were attached to the lens of some of our CCD video cameras. These gratings had 500 or 1000 grooves per mm, depending on the device. The emission spectra recorded in the framework of this survey have been analyzed with the CHIMET software, which is able to identify and measure emission lines in these signals (Madiedo et al. 2013b).

## 3. RESULTS

Variable weather conditions were found each year. Thus, bad weather over Spain made data acquisition impossible for SPMN stations during January 1-3, 2008. That year our first data were collected during January 3-4 from two stations located in Andalusia: Sevilla and the mobile system at Cerro Negro. The mobile system was setup during the afternoon of January 3 in order to provide a double-station system coordinated with our station at Seville during the period of maximum activity of the Quadrantid shower, which according to the information provided by the International Meteor Organization (IMO) was predicted to peak on January 4 at 6h40m UTC (www.imo.net). Nevertheless, weather conditions were completely unfavourable during the first hours after sunset, since the sky was completely cloudy in Cerro Negro and scattered showers were present also in the Seville area. Despite this, the devices at the mobile station remained ready for operation and covered with sheets of plastic to protect them from the rain, in order to make profit of an eventual improvement in weather conditions. Fortunately this improvement took place around 2h10m UTC and the Quadrantids could be observed during the predicted peak, together with a number of transient luminous events (TLE) in the atmosphere (sprites) that were also recorded.

In January 2009, bad weather conditions did not allow obtaining any double station Quadrantid meteors, despite our meteor network was expanded and four additional observing stations were available by that time. Thus, just single station Quadrantid trails were imaged, and so no orbital data could be inferred from these. In January 2010, unfavourable weather interfered with the observation of this shower again, and just stations #8, 9 and 10, located in the north of Spain, could image some double-station Quadrantids. The situation, however, was far more favourable for our stations in the south of Spain during January 1-7, 2011 and 2012. Thus, the night sky was completely clear and transparent during most of that period, and the Quadrantids could be observed in that area of the country. Meteor rates measured by the forward-scatter device during January 3-4, 2011 are shown in Figure 1, where the counts registered versus time are plotted. This graph shows that the maximum activity was registered between 0h and 06 h UTC on January 4, which is consistent with the results obtained by IMO (www.imo.net/quadrantids2011). Figure 1 also exhibits two maxima, which also agree with the results published by IMO. This device was not operated during the rest of the years considered in this work.

In total, from 2008 to 2012, over 300 double-station Quadrantids were imaged. The brightest of these was a mag. -7.5 ± 0.5 fireball recorded on 4 January 2011 at 2h32m37.0 ± 0.1s from stations #2 and 7 in Table 1 (Cerro Negro and El Arenosillo, respectively). We have only considered here those events with a convergence angle Q higher than 20º in order to obtain more accurate orbital elements. This is the angle between the two planes delimited by the observing sites and the meteor path in the triangulation, and it measures the quality of the determination of the atmospheric trajectory (Ceplecha 1987). These events are listed in Table 2, where we have included the absolute magnitude (M), the initial (preatmospheric) photometric mass of the meteoroid ($m_p$), the beginning and ending heights ($H_b$ and $H_e$, respectively), the position (J2000.0) of the geocentric radiant ($\alpha_g$, $\delta_g$), and the preatmospheric ($V_\infty$), geocentric ($V_g$) and heliocentric velocities ($V_h$). A code has been assigned to each event for identification with the format QDYYEE, where Q indicates that the meteor belongs to the Quadrantid stream, D is the day of the month (which ranges between 2 and 4 for the meteors analyzed here), and YY the last two digits of the recording year. The two digits EE are employed to number meteors recorded during the same night and considered in this analysis, so that 00 is assigned to the first meteor imaged, 01 to the next one and so on. From the radiant position, apparition time and velocities estimated for these meteors we have derived the orbital elements shown in Table 3.

The video spectrographs obtained four good quality emission spectra produced by Quadrantid meteors with magnitudes ranging from -4.7 to -7.0. These spectra correspond to events Q41110, Q41112, Q41221, and Q41238 in Table 2. Seven additional spectra produced by less luminous meteors were also recorded, but unfortunately the signal was too faint and so these spectra were not taken into consideration for this research. We have employed the CHIMET software to process these spectra, which follows the analysis procedure described in Madiedo et al. (2013b). Thus, for each spectrum recorded by the spectrographs we first deinterlaced the corresponding video file. Next, the video frames containing the emission spectrum were dark-frame subtracted and flat-fielded. Then, we tried to identify typical lines appearing in meteor spectra (Ca, Fe, Mg, and Na multiplets) to calibrate the signal in wavelengths. Once this calibration was performed, the intensity of the signal was corrected by taking into account the spectral efficiency of the recording instrument. The results are shown in Figures 2 to 5, where multiplet numbers are given according to Moore (1945). The line of atmospheric O I at 777.4 nm is very prominent. The most important emissions associated with the meteoroid composition correspond to the H and K lines of ionized

Ca, at 393.3 and 396.8 nm respectively. These Ca II-1 lines appear blended in the signal. The contributions of several Fe I multiplets have been also identified, together with the emission of the Mg I-2 triplet at 517.3 nm, and the Na I-1 doublet at 588.9 nm.

## 4. DISCUSSION

### 4.1. Meteor initial and final heights

The beginning and final heights of events listed in Table 2 are shown in Figures 6 and 7, respectively. As previously found by other authors (Koten et al. 2006, Jenniskens 2004), the beginning height $H_b$ of Quadrantid meteors was below 110 km above the sea level, lower than the value found for cometary showers such as the Leonids, the Orionids and the Perseids. Besides, this parameter was found to increase with increasing meteoroid mass. The dependence of $H_b$ on the logarithm of the photometric mass was described by means of a linear relationship (solid line in Figure 6). The slope of this line yields 0.85 ± 0.65, which means that the increase of the beginning height with mass is less pronounced for the Quadrantids than for members of cometary showers such as the Leonids (a = 9.9 ± 1.5), the Perseids (a = 7.9 ± 1.3), the Taurids (a = 6.6 ± 2.2) and the Orionids (a = 5.02 ± 0.65), but more pronounced than for the Geminids (a = 0.46 ± 0.26) (Koten et al. 2004), which have an asteroidal origin (Jenniskens 2004). The slope obtained for the Quadrantids is similar to the 1.1 ± 0.5 value obtained for the ρ-Geminids, which are produced by tough cometary materials (Madiedo 2015). This is consistent with the result obtained by Koten et al. (2006), who proposed that the parent body of the Quadrantids is a dormant comet. Nevertheless, the slope we have obtained for the line that describes the dependence of $H_b$ with the logarithm of the photometric mass for our 85 Quadrantid events is significantly lower than the 3.4 ± 0.8 value obtained by Koten et al. (2006) for 44 Quadrantids. Thus, our results for the beginning height for the Quadrantids are in much better agreement with the values derived by Jenniskens (2004). Contrary to the results shown in Figure 1 in Koten et al. (2006), we have observed a significant number of events with $H_b$ below 95 km for values of the logarithm of the photometric mass ranging between 0 and -2. This agrees with the results shown in Figure 5 in Jenniskens (2004).

As expected, the terminal point of the luminous trajectory of the Quadrantids occurs at lower altitudes as the meteoroid mass increases (Figure 7). This behaviour can be also described by means of a linear relationship between $H_e$ and the logarithm of the photometric mass (solid line in Figure 7). The slope of this line yields -4.3 ± 0.6. Our results show that the Quadrantids, with a preatmospheric velocity of about 43 km $s^{-1}$, do not penetrate as deep as the Geminids (Jenniskens 2004, Koten et al. 2006), which have a preatmospheric velocity of ~36 km $s^{-1}$. However, they penetrate deeper than the Perseids, which have a cometary origin and a preatmospheric velocity of ~61 km $s^{-1}$.

## 4.2. Tensile strength

Almost all of the multi-station Quadrantid meteors recorded by us are characterized by smooth light curves. However, three double-station fireballs exhibited a flare along their atmospheric path. These were the mag -7.0 ± 0.5 bolide imaged on 2 January 2010 at 5h10m09.1 ± 0.1 s UTC, the mag. -6.5 ± 0.5 event recorded on 4 January 2011 at 1h53m58.8 ± 0.1 s UTC, and the mag. -5.0 ± 0.5 fireball imaged on 4 January 2011 at 6h05m37.6 ± 0.1 s UTC. The height and velocity at which their flares took place are shown in Table 4. These flares are typically produced when meteoroids break-up as these particles penetrate denser atmospheric regions. Thus, once the overloading pressure becomes larger than the particle strength, the meteoroid experiences fragmentation. Immediately after this takes place, a flare is produced as a consequence of the fast ablation of the tiny fragments delivered to the thermal wave in the fireball's bow shock. The aerodynamic pressure under which these flares take place can be used as a proxy for an estimation of the strength of the particle (Trigo-Rodríguez & Llorca 2006). The aerodynamic pressure S at which this break-up took place can be calculated from the following relationship (Bronshten 1981):

$$S = \rho_{atm} \cdot v^2 \qquad (1)$$

where v is the velocity of the meteoroid at the disruption point and $\rho_{atm}$ the atmospheric density at the height where this fracture takes place. We have calculated the atmospheric density by using the US standard atmosphere model (U.S. Standard Atmosphere 1976). In this way, we infer that the progenitor meteoroids of the above-mentioned fireballs exhibited their flares under the dynamic pressures shown in Table 4. These values are similar to the average tensile strength found by Trigo-Rodríguez & Llorca (2006, 2007) for meteoroids in the Quadrantid stream ( ~ $2 \cdot 10^5$ dyn $cm^{-2}$), except for the fireball recorded on 2 January 2010. Thus, the tensile strength estimated for this meteoroid is one order of magnitude below.

### 4.3. Emission spectra

To obtain insight into the chemical nature of Quadrantid meteoroids we have analyzed the relative intensity of the Na I-1, Mg I-2 and Fe I-15 multiplets (Borovička et al. 2005). For this purpose, for each video spectrum the intensity of the corresponding emission lines was measured frame by frame and then corrected for the instrumental efficiency. The contributions in each frame were then added to obtain the integrated intensity for each line along the meteor path. The ternary diagram in Figure 8 shows the relative intensity of the emission from the Na I-1, Mg I-2 and Fe I-15 multiplets. The solid curve in this diagram represents the expected relative intensity, as a function of meteor velocity, for chondritic meteoroids (Borovička et al. (2005)). The positions in this plot describing the four spectra presented in this work show that Quadrantid meteoroids, with a velocity of ~ 43 km $s^{-1}$, can be regarded as normal, in the sense defined in (Borovička et al. (2005). However, these also reveal that the progenitor meteoroids have suffered a loss of Na in different degrees. This depletion in volatiles in Quadrantid meteoroids was previously reported by Koten et al. (2006), who proposed that this would be the result that the commonly accepted parent body of the Quadrantid stream (NEO 2003 EH1) is a dormant comet. The scenario presented by Neslušan et al. (2013), who proposed that Comet 96P/Machholz and asteroid 2003 EH1 could be both the parent bodies of this stream, is also compatible with this result. Thus, according to this hypothesis, the Quadrantid stream includes two different populations of meteoroids originated from 2003 EH1 and Comet 96P, and since both objects had a common progenitor NEO 2003 EH1 would be a cometary body.

## 5. CONCLUSIONS

We have analyzed the activity of the Quadrantid meteor shower from 2008 to 2012. From this survey 85 orbits have been obtained for multi-station meteors providing values of the convergence angle Q higher than 20 degrees. The beginning height for these meteors, which increases with increasing meteoroid mass, was below 110 km above the sea level, lower than the value found for cometary showers such as the Leonids, the Orionids and the Perseids. The dependence of this beginning height on the logarithm of the photometric mass suggests that Quadrantid meteors are produced by tough cometary materials. Most meteors exhibited a quasy-continuous ablation behaviour, with smooth lightcurves. The tensile strength estimated for Quadrantid meteoroids exhibiting flares during their atmospheric path ranges between The tensile strength estimated for Quadrantid meteoroids ranges from (2.7 ± 0.7)·$10^4$ to (3.5 ± 0.3)·$10^5$ dyn $cm^{-2}$. Besides, the emission spectra of four Quadrantid meteors with absolute magnitudes ranging between -4.7 and -7.0 have been analyzed. According to these spectra, Quadrantid meteoroids can be regarded as normal. These signals also reveal different degrees of Na depletion among these particles. Our results indicate that the Quadrantids appear to be cometary in nature, which suggests that the commonly accepted parent body of this stream, NEO 2003 EH1, is a comet fragment.

**ACKNOWLEDGEMENTS**

The meteor observing stations at Sevilla, Huelva, Cerro Negro, El Arenosillo, Sierra Nevada and La Hita have been funded by the first author. We thank AstroHita Foundation for its continuous support in the operation of the meteor observing station located at La Hita Astronomical Observatory.

**REFERENCES**.

# FIGURES

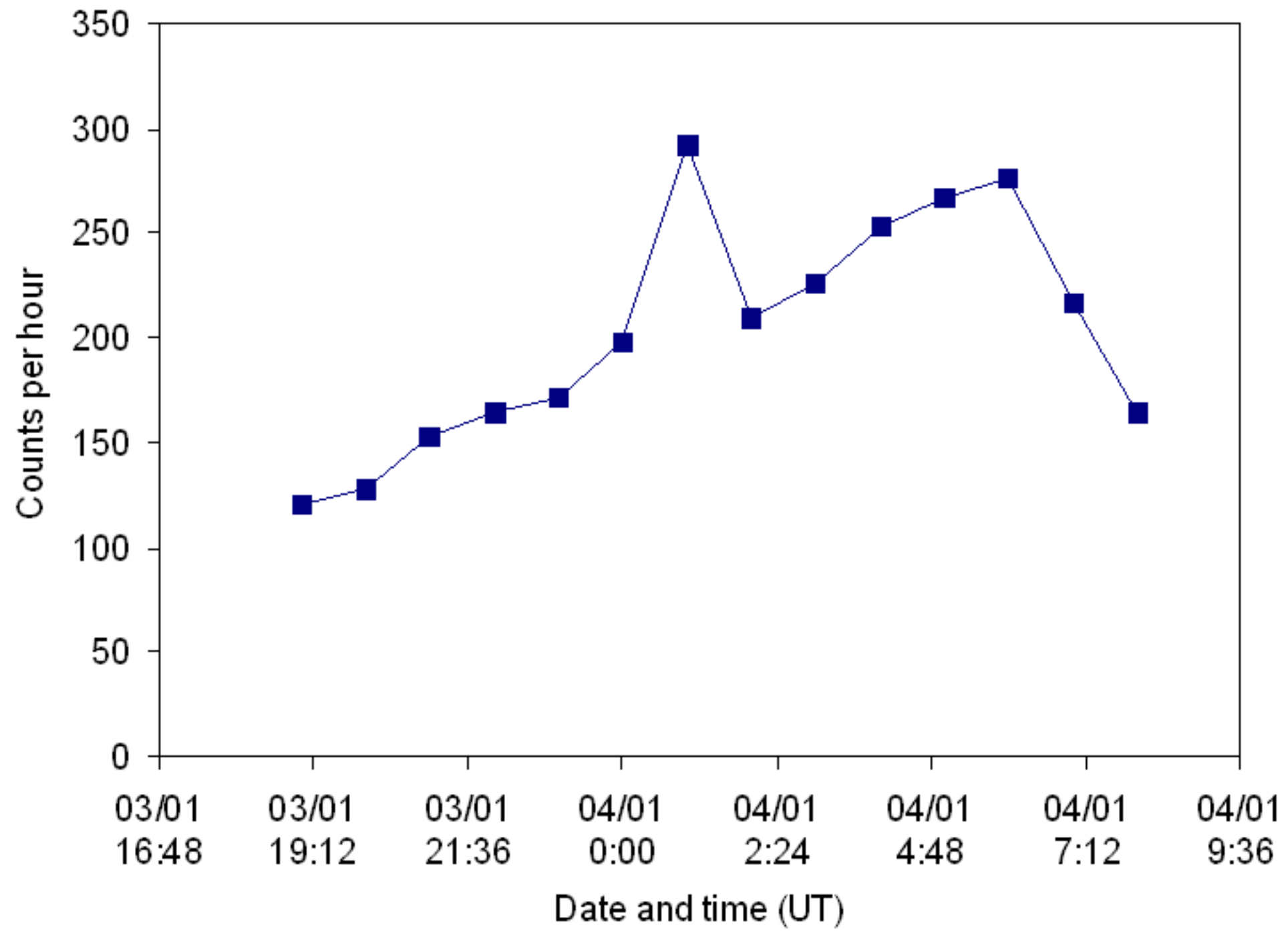

Figure 1. Activity level obtained by the forward scatter systems operating on Jan. 3-4 2011 at the frequency of 143.05 MHz from Madrid.

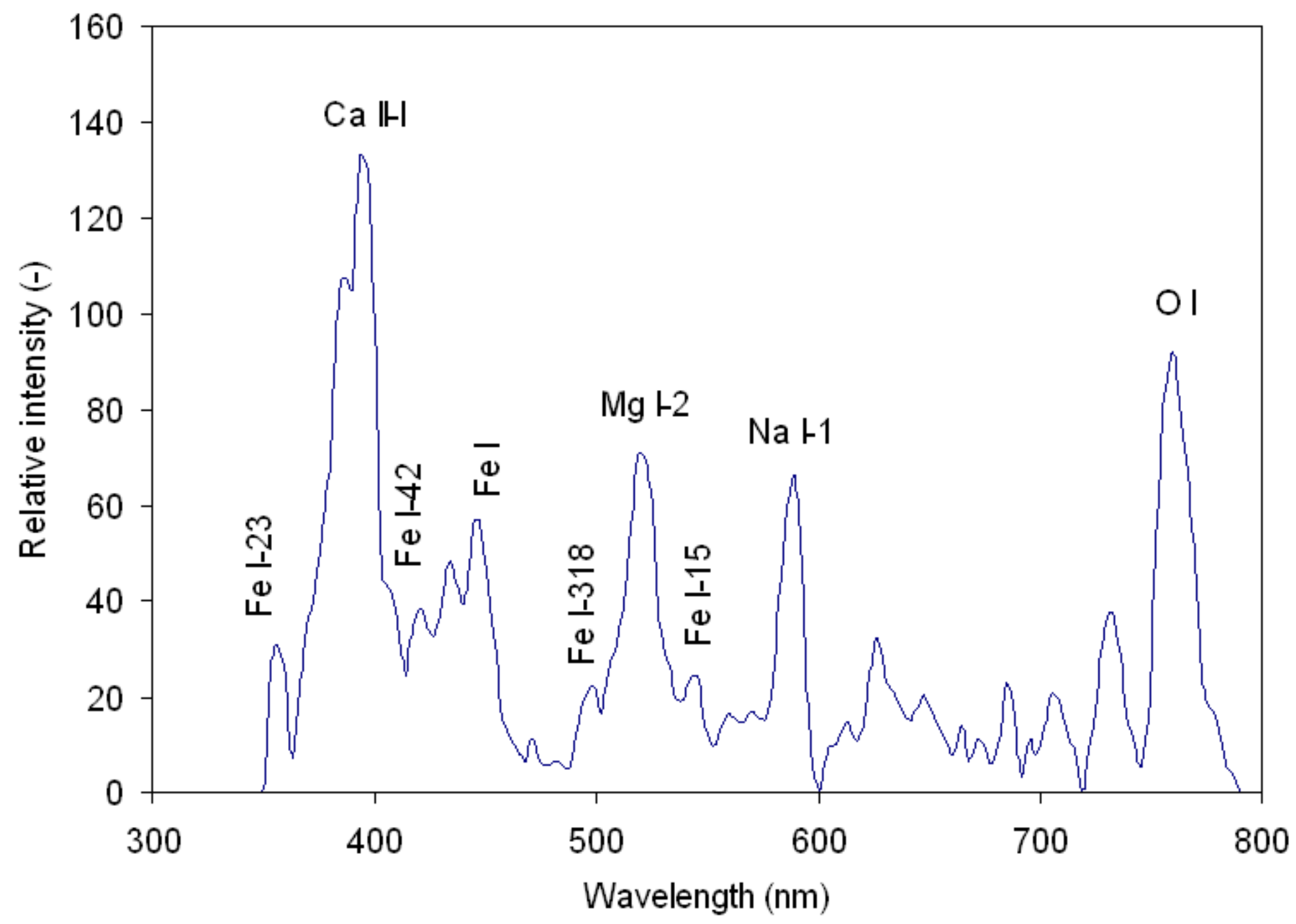

Figure 2. Calibrated emission spectrum of the Q41110 meteor.

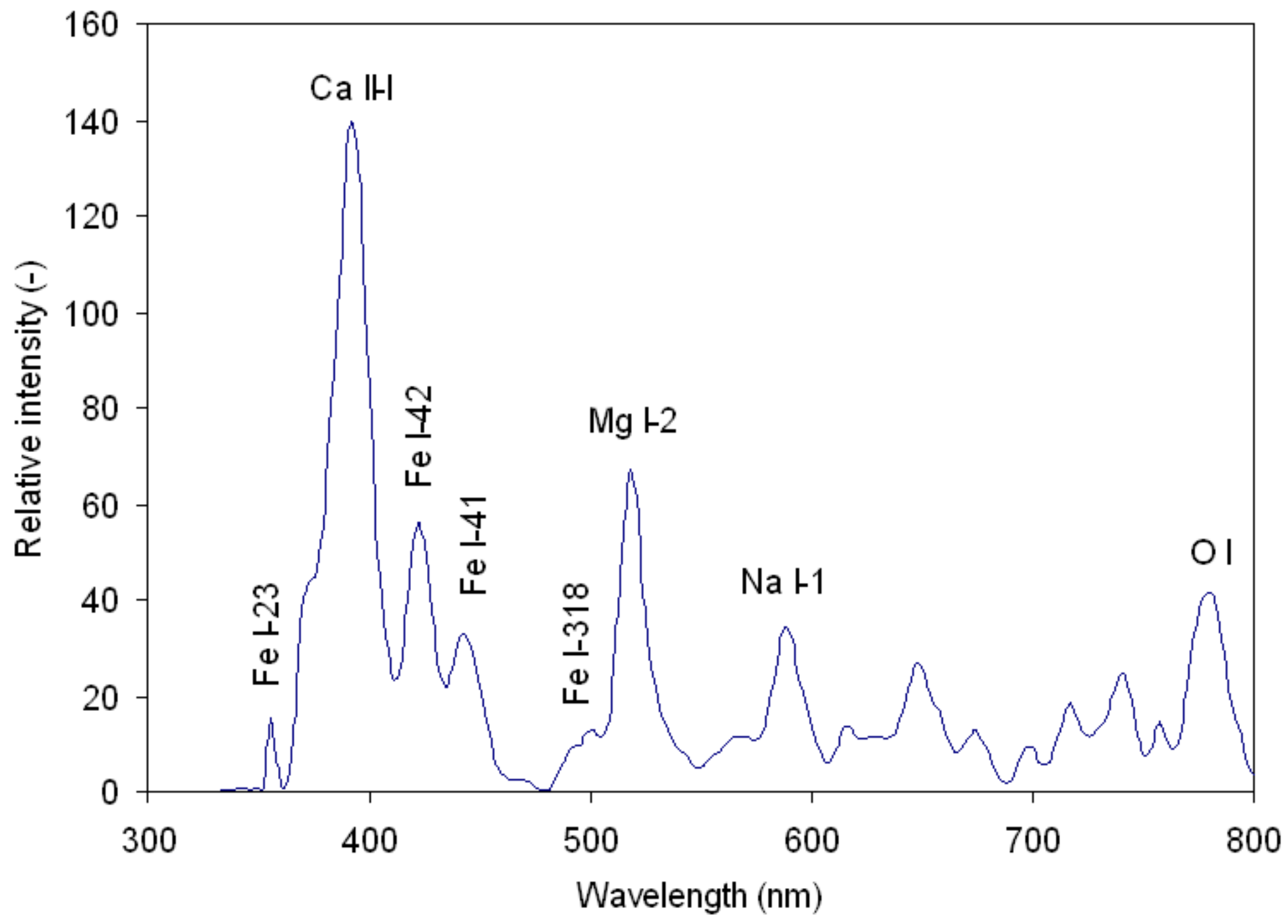

Figure 3. Calibrated emission spectrum of the Q41112 meteor.

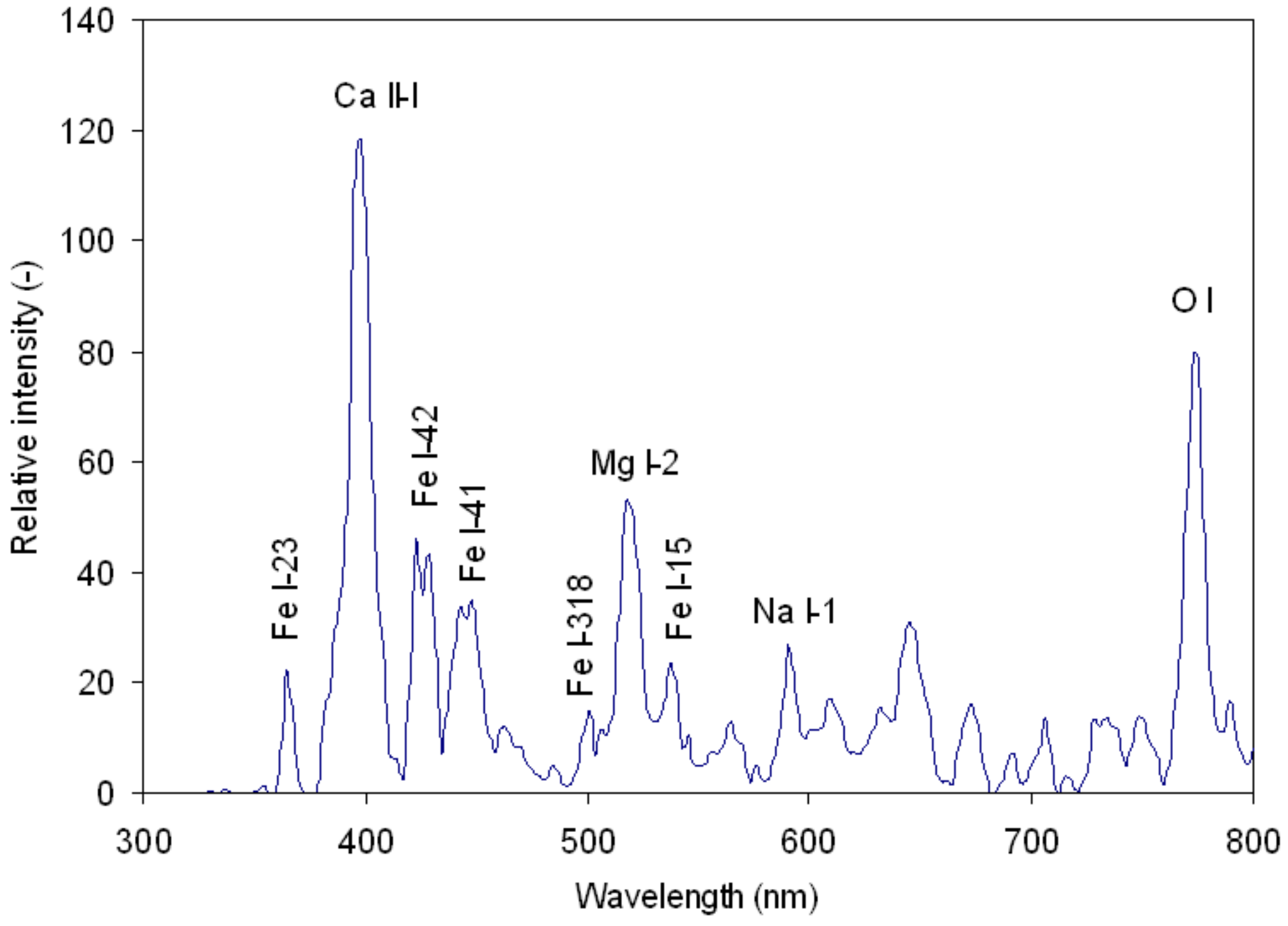

Figure 4. Calibrated emission spectrum of the Q41221 meteor.

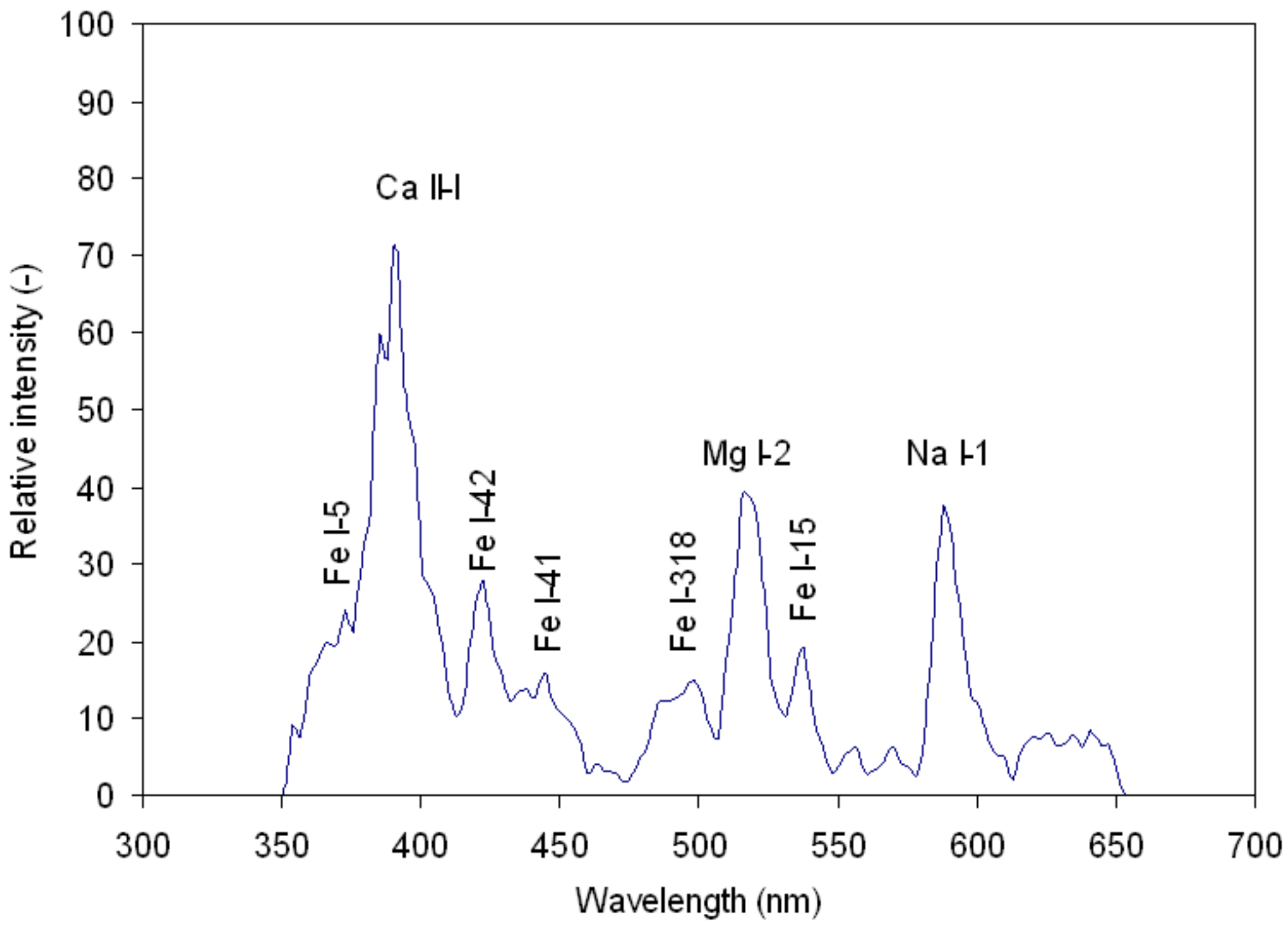

Figure 5. Calibrated emission spectrum of the Q41238 meteor.

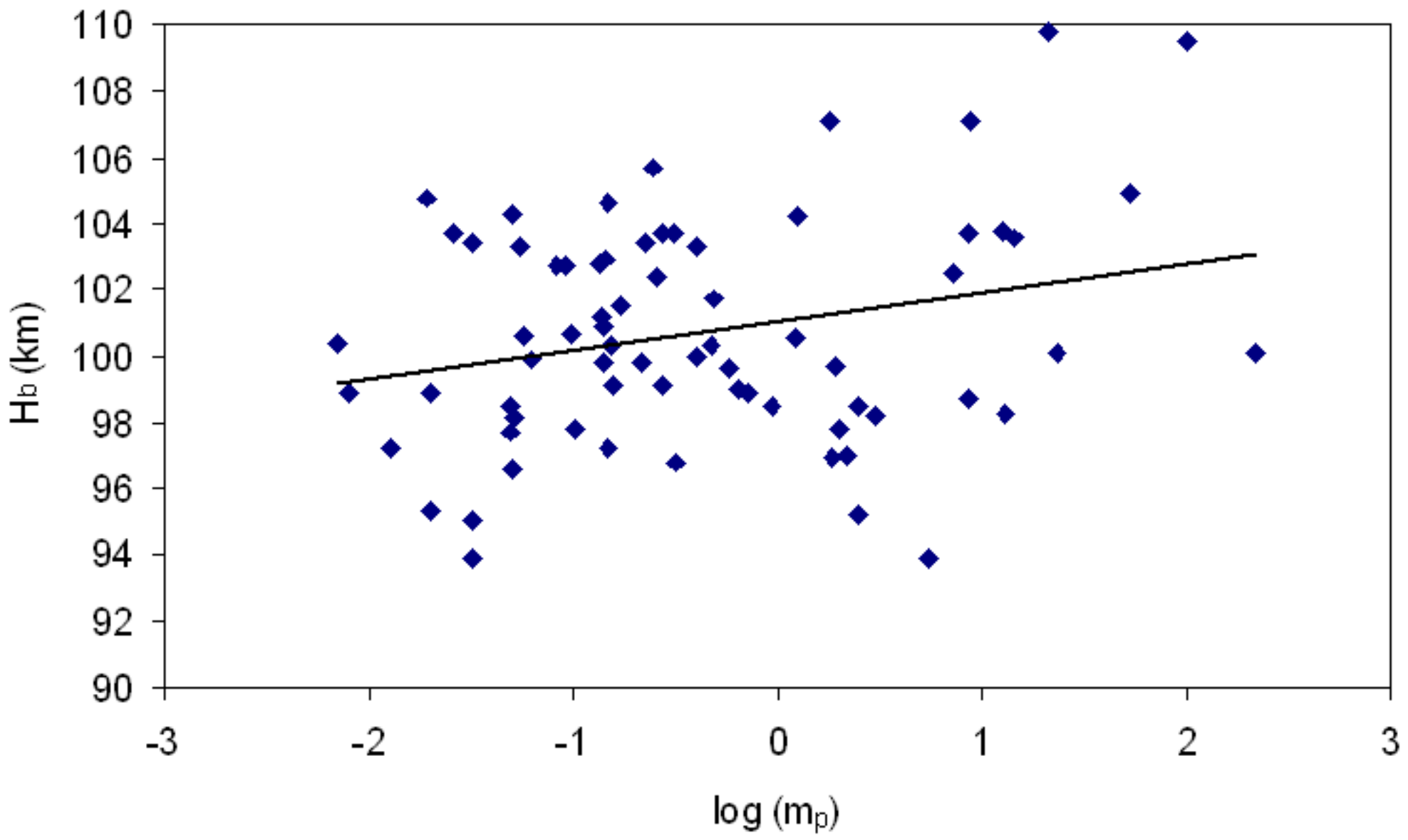

Figure 6. Meteor beginning height $H_b$ vs. logarithm of the photometric mass $m_p$ of the meteoroid. Solid line: linear fit for measured data.

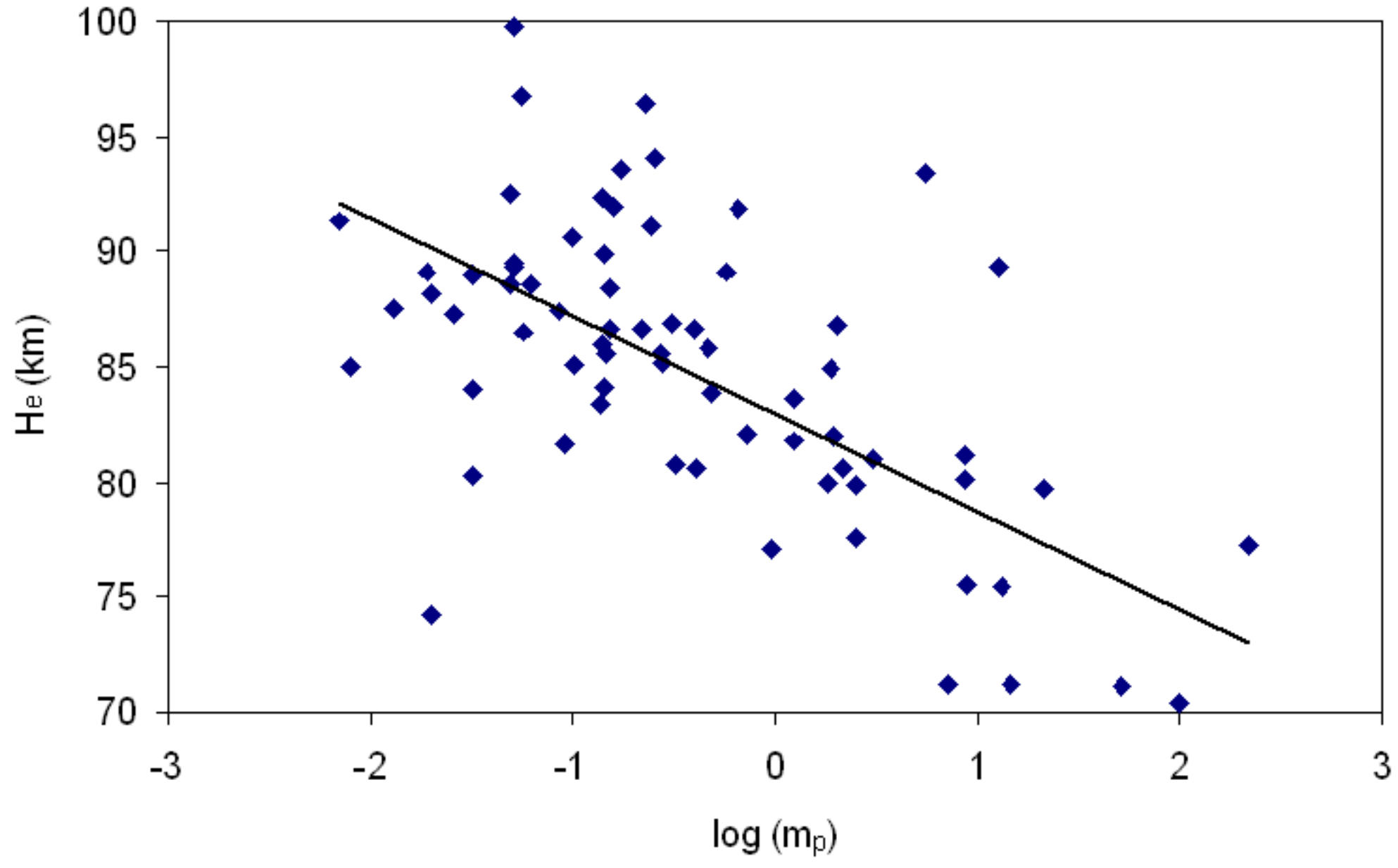

Figure 7. Meteor terminal height $H_e$ vs. logarithm of the photometric mass $m_p$ of the meteoroid. Solid line: linear fit for measured data.

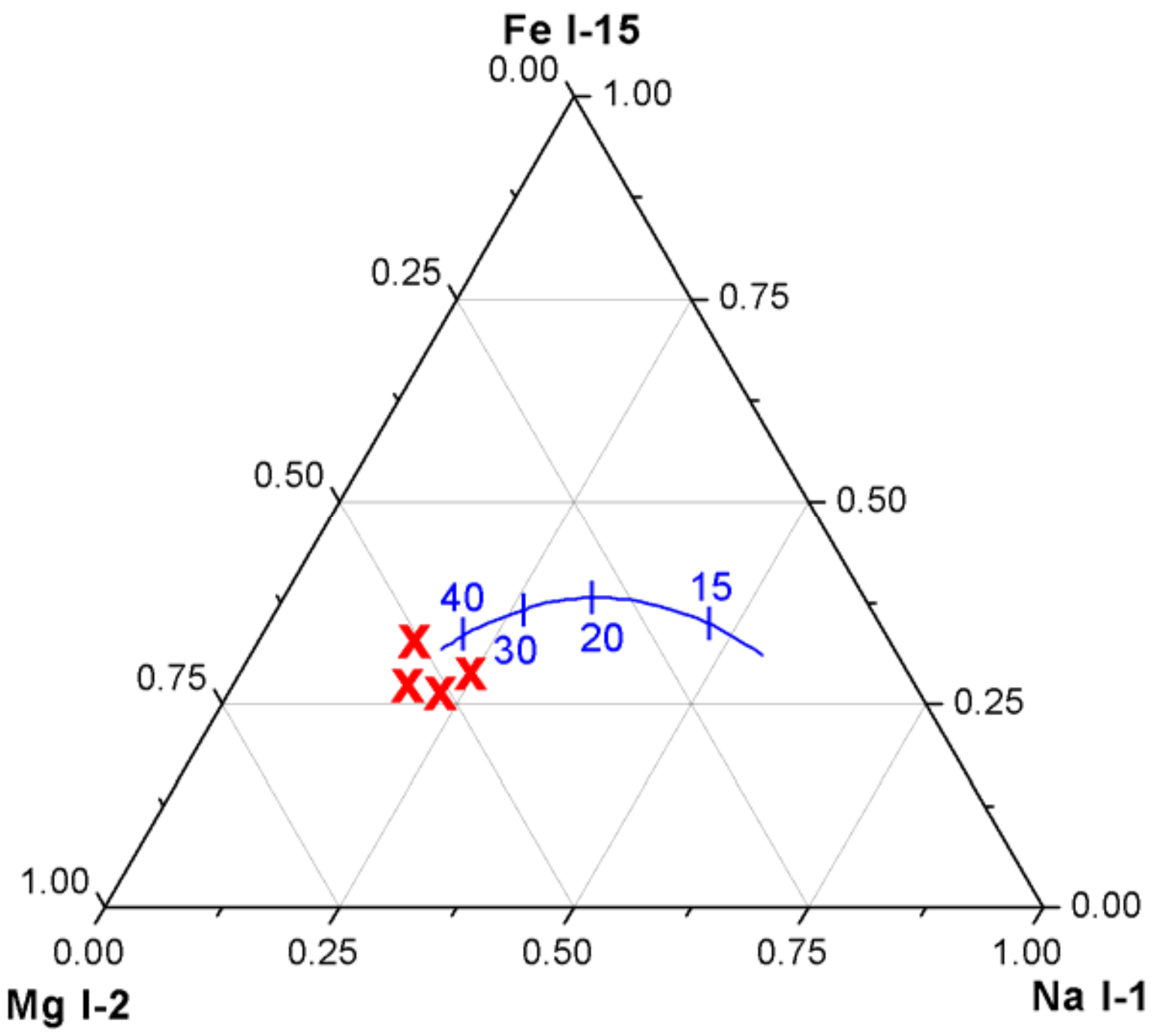

Figure 8. Expected relative intensity (solid line), as a function of meteor velocity (in km $s^{-1}$), of the Na I-1, Mg I-2 and Fe I-15 multiplets for chondritic meteoroids (Borovička et al., 2005). The crosses shows the experimental relative intensities obtained for the Quadrantid spectra discussed in the text.

## TABLES

Table 1. Geographical coordinates of the SPMN meteor observing stations involved in this work.

| Station # | Station name | Longitude | Latitude (N) | Alt. (m) |
|---|---|---|---|---|
| 1 | Sevilla | 5º 58′ 50" W | 37º 20′ 46" | 28 |
| 2 | Cerro Negro | 6º 19′ 35" W | 37º 40′ 19" | 470 |
| 3 | La Hita | 3º 11' 00" W | 39º 34' 06" | 674 |
| 4 | Huelva | 6º 56' 11" W | 37º 15' 10" | 25 |
| 5 | La Murta | 1º 12' 10" W | 37º 50' 25" | 400 |
| 6 | Sierra Nevada | 3º 23′ 05" W | 37º 03′ 51" | 2896 |
| 7 | El Arenosillo | 6º 43′ 58" W | 37º 06′ 16" | 40 |
| 8 | Folgueroles | 2º 19' 33" E | 41º 56' 31" | 580 |
| 9 | OAdM (Montsec) | 0º 43' 46" E | 42º 03' 05" | 1570 |
| 10 | Montseny | 2º 31' 14" E | 41º 43' 17" | 300 |

Table 2. Trajectory and radiant data for 85 multi-station Quadrantid meteors recorded from 2008 to 2012 (J2000).

| Meteor Code | Date | Time (UT) ±0.1s | M ±0.5 | $m_p$ (g) | $H_b$ (km) ±0.5 | $H_e$ (km) ±0.5 | $\alpha_g$ (º) | $\delta_g$ (º) | $V_\infty$ (km s$^{-1}$) | $V_g$ (km s$^{-1}$) | $V_h$ (km s$^{-1}$) |
|---|---|---|---|---|---|---|---|---|---|---|---|
| Q40800 | Jan. 4, 2008 | 2h32m54.4s | 2.0 | 0.019±0.002 | 104.8 | 89.1 | 228.6±0.1 | 48.3±0.1 | 43.7±0.5 | 42.0±0.5 | 38.8±0.5 |
| Q40801 | Jan. 4, 2008 | 2h34m39.4s | 0.0 | 0.152±0.002 | 100.3 | 88.4 | 231.2±0.1 | 48.9±0.1 | 43.0±0.7 | 41.3±0.7 | 39.1±0.7 |
| Q40802 | Jan. 4, 2008 | 2h38m23.0s | 1.0 | 0.051±0.005 | 109.3 | 99.8 | 234.3±0.3 | 44.9±0.2 | 43.0±0.5 | 41.3±0.5 | 38.5±0.5 |
| Q40803 | Jan. 4, 2008 | 2h43m03.2s | -4.0 | 8.62±0.91 | 103.7 | 81.2 | 231.6±0.1 | 48.3±0.1 | 42.7±0.4 | 40.9±0.4 | 38.7±0.4 |
| Q40804 | Jan. 4, 2008 | 2h55m59.9s | -2.5 | 1.92±0.21 | 99.7 | 82.0 | 233.2±0.2 | 48.2±0.1 | 42.7±0.5 | 40.9±0.5 | 39.0±0.5 |
| Q40805 | Jan. 4, 2008 | 2h59m05.7s | 0.0 | 0.149±0.021 | 104.6 | 86.6 | 227.4±0.3 | 49.2±0.2 | 43.7±0.6 | 42.1±0.6 | 38.9±0.6 |
| Q40806 | Jan. 4, 2008 | 3h10m04.8s | 1.0 | 0.057±0.006 | 100.6 | 86.5 | 230.4±0.1 | 48.5±0.1 | 43.2±0.5 | 41.5±0.5 | 38.8±0.5 |
| Q40807 | Jan. 4, 2008 | 3h21m35.1s | 0.0 | 0.135±0.004 | 107.8 | 83.4 | 227.1±0.3 | 47.2±0.2 | 43.9±0.6 | 42.2±0.6 | 38.4±0.6 |
| Q40808 | Jan. 4, 2008 | 3h25m08.5s | 2.5 | 0.013±0.005 | 97.2 | 87.5 | 229.9±0.2 | 49.3±0.1 | 42.5±0.5 | 40.7±0.5 | 38.5±0.5 |
| Q40809 | Jan. 4, 2008 | 3h27m37.2s | 1.5 | 0.032±0.004 | 103.4 | 80.3 | 228.3±0.1 | 49.8±0.1 | 43.5±0.5 | 41.8±0.5 | 39.1±0.5 |
| Q40810 | Jan. 4, 2008 | 3h35m42.3s | 0.5 | 0.084±0.009 | 102.7 | 87.4 | 228.2±0.2 | 48.1±0.1 | 43.9±0.5 | 42.2±0.5 | 38.7±0.5 |
| Q40811 | Jan. 4, 2008 | 3h42m37.7s | -2.0 | 1.23±0.15 | 100.5 | 83.6 | 231.4±0.1 | 49.3±0.1 | 42.2±0.5 | 40.4±0.5 | 38.6±0.5 |
| Q40812 | Jan. 4, 2008 | 3h47m21.8s | 0.5 | 0.092±0.011 | 102.7 | 81.7 | 230.3±0.1 | 48.4±0.1 | 42.8±0.6 | 41.1±0.6 | 38.5±0.6 |
| Q40813 | Jan. 4, 2008 | 3h50m11.2s | 3.0 | 0.007±0.001 | 100.4 | 91.3 | 229.9±0.2 | 49.1±0.1 | 43.1±0.4 | 41.4±0.4 | 38.8±0.4 |
| Q40814 | Jan. 4, 2008 | 3h51m21.6s | 0.0 | 0.143±0.016 | 102.9 | 84.1 | 229.9±0.2 | 48.3±0.1 | 43.5±0.4 | 41.8±0.4 | 38.8±0.4 |
| Q40815 | Jan. 4, 2008 | 3h52m51.0s | -0.5 | 2.03±0.27 | 97.8 | 86.8 | 233.9±0.1 | 47.9±0.1 | 42.2±0.5 | 40.4±0.5 | 38.7±0.5 |
| Q40816 | Jan. 4, 2008 | 3h59m54.8s | 1.5 | 0.032±0.003 | 95.0 | 89.0 | 231.2±0.1 | 48.9±0.1 | 42.6±0.5 | 40.8±0.5 | 38.7±0.5 |
| Q40817 | Jan. 4, 2008 | 4h00m08.1s | 0.5 | 0.102±0.012 | 97.8 | 85.1 | 229.5±0.1 | 50.2±0.1 | 42.5±0.6 | 40.7±0.6 | 38.8±0.6 |
| Q40818 | Jan. 4, 2008 | 4h02m43.1s | 1.0 | 0.051±0.006 | 96.6 | 89.5 | 230.8±0.1 | 47.2±0.1 | 43.3±0.6 | 41.6±0.6 | 38.5±0.6 |
| Q40819 | Jan. 4, 2008 | 4h07m20.9s | 2.0 | 0.020±0.002 | 98.9 | 88.2 | 231.5±0.2 | 47.5±0.1 | 43.5±0.5 | 41.8±0.5 | 39.0±0.5 |
| Q40820 | Jan. 4, 2008 | 4h09m58.4s | -2.5 | 1.82±0.20 | 107.1 | 80.0 | 230.9±0.2 | 49.2±0.1 | 43.1±0.6 | 41.4±0.6 | 39.2±0.6 |
| Q21000 | Jan. 2,2010 | 5h10m09.1s | -5.0 | 22±3 | 97.1 | 84.2 | 223.7±0.4 | 49.2±0.4 | 44.1±0.7 | 42.5±0.7 | 38.2±0.7 |
| Q31100 | Jan. 3, 2011 | 18h38m03.5s | -1.5 | 0.652±0.061 | 99.0 | 91.8 | 228.5±0.1 | 48.4±0.1 | 43.1±0.5 | 41.7±0.5 | 38.5±0.5 |
| Q31101 | Jan. 3, 2011 | 22h20m05.6s | -3.5 | 5.52±0.70 | 93.9 | 93.5 | 233.1±0.1 | 48.4±0.1 | 42.0±0.5 | 40.4±0.5 | 38.7±0.5 |
| Q31102 | Jan. 3, 2011 | 23h26m57.5s | -3.0 | 2.7±0.3 | 101.2 | 85.0 | 230.7±0.5 | 50.7±0.5 | 43.2±0.4 | 41.9±0.4 | 40.0±0.4 |
| Q31103 | Jan. 3, 2011 | 23h35m41.5s | -4.0 | 8.02±0.61 | 111.2 | 88.5 | 227.7±0.5 | 49.5±0.5 | 42.1±0.5 | 40.4±0.5 | 37.8±0.5 |
| Q41100 | Jan. 4, 2011 | 0h05m2.6s | -2.5 | 1.86±0.16 | 96.9 | 84.9 | 231.3±0.1 | 48.5±0.1 | 42.9±0.5 | 41.2±0.5 | 38.9±0.5 |
| Q41101 | Jan. 4, 2011 | 0h13m08.3s | -1.0 | 0.393±0.05 | 103.2 | 86.3 | 219.8±0.4 | 48.2±0.4 | 44.3±0.5 | 42.6±0.5 | 37.3±0.5 |
| Q41102 | Jan. 4, 2011 | 0h22m19.5s | 0.0 | 0.171±0.021 | 101.5 | 93.6 | 233.7±0.3 | 50.8±0.2 | 41.6±0.6 | 39.9±0.6 | 39.0±0.6 |
| Q41103 | Jan. 4, 2011 | 0h27m15.4s | -1.0 | 0.379±0.05 | 107.5 | 93.4 | 230.1±0.3 | 51.2±0.2 | 41.6±0.4 | 39.9±0.4 | 38.6±0.4 |
| Q41104 | Jan. 4, 2011 | 0h28m15.7s | 2.0 | 0.033±0.005 | 99.5 | 89.6 | 230.2±0.3 | 50.3±0.2 | 41.5±0.3 | 39.8±0.3 | 38.2±0.3 |
| Q41105 | Jan. 4, 2011 | 0h48m14.2s | -5.0 | 26±4 | 105.8 | 86.1 | 227.4±0.5 | 51.5±0.5 | 43.2±0.5 | 41.5±0.5 | 39.3±0.5 |
| Q41106 | Jan. 4, 2011 | 0h59m42.4s | -2.0 | 1.18±0.14 | 96.6 | 86.3 | 228.1±0.5 | 50.8±0.5 | 43.0±0.5 | 41.3±0.5 | 39.0±0.5 |
| Q41107 | Jan. 4, 2011 | 1h38m12.1s | 0.0 | 0.201±0.021 | 99.0 | 88.8 | 229.1±0.3 | 49.1±0.2 | 42.5±0.3 | 40.8±0.3 | 38.3±0.3 |
| Q41108 | Jan. 4, 2011 | 1h43m49.2s | 1.5 | 0.032±0.003 | 93.9 | 84.0 | 229.6±0.3 | 47.8±0.1 | 43.8±0.4 | 41.1±0.4 | 38.8±0.4 |
| Q41109 | Jan. 4, 2011 | 1h44m42.6s | -4.0 | 9.77±0.98 | 104.2 | 83.4 | 227.9±0.3 | 52.7±0.1 | 42.4±0.3 | 40.7±0.3 | 39.5±0.3 |
| Q41110 | Jan. 4, 2011 | 1h53m58.8s | -6.5 | 100±11 | 108.0 | 70.4 | 228.0±0.3 | 49.4±0.2 | 43.3±0.5 | 41.6±0.5 | 38.7±0.5 |
| Q41111 | Jan. 4, 2011 | 1h58m16.5s | -4.5 | 14.5±1.6 | 103.6 | 71.2 | 227.7±0.4 | 50.9±0.2 | 42.6±0.6 | 40.8±0.6 | 38.7±0.6 |
| Q41112 | Jan. 4, 2011 | 2h32m37.0s | -7.5 | 220±23 | 100.1 | 77.2 | 233.1±0.2 | 48.2±0.1 | 42.3±0.5 | 40.5±0.5 | 39.0±0.5 |
| Q41113 | Jan. 4, 2011 | 3h07m43.0s | 0.0 | 0.143±0.018 | 100.8 | 84.0 | 229.6±0.2 | 50.8±0.1 | 42.8±0.3 | 41.1±0.3 | 39.2±0.3 |
| Q41114 | Jan. 4, 2011 | 3h11m31.5s | -0.5 | 0.210±0.003 | 99.8 | 84.8 | 230.0±0.3 | 52.1±0.3 | 42.1±0.3 | 40.4±0.3 | 39.4±0.3 |
| Q41115 | Jan. 4, 2011 | 3h12m47.4s | 1.0 | 0.055±0.004 | 103.3 | 96.8 | 231.7±0.1 | 48.2±0.1 | 43.1±0.6 | 41.3±0.6 | 38.9±0.6 |
| Q41116 | Jan. 4, 2011 | 4h23m00.9s | -1.0 | 0.471±0.056 | 100.3 | 85.8 | 234.9±0.1 | 49.9±0.1 | 41.7±0.7 | 40.0±0.7 | 39.3±0.7 |
| Q41117 | Jan. 4, 2011 | 5h43m14.5s | -5.0 | 30±5 | 105.8 | 85.4 | 230.1±0.3 | 52.1±0.2 | 42.0±0.3 | 40.3±0.3 | 39.2±0.3 |
| Q41118 | Jan. 4, 2011 | 6h05m37.6s | -7.0 | 109±12 | 101.8 | 67.1 | 229.9±0.5 | 49.9±0.5 | 41.2±0.5 | 39.5±0.5 | 37.8±0.5 |
| Q41200 | Jan. 4, 2012 | 2h09m32.1s | -4.5 | 12.8±1.4 | 103.8 | 89.3 | 230.1±0.1 | 48.6±0.1 | 43.6±0.7 | 41.9±0.7 | 39.1±0.7 |
| Q41201 | Jan. 4, 2012 | 2h25m44.0s | -1.0 | 0.403±0.05 | 100.0 | 86.6 | 231.4±0.1 | 48.0±0.1 | 43.2±0.6 | 41.5±0.6 | 38.9±0.6 |
| Q41202 | Jan. 4, 2012 | 3h40m26.4s | -2.0 | 1.25±0.15 | 104.2 | 81.8 | 233.1±0.3 | 49.7±0.2 | 42.0±0.6 | 40.2±0.6 | 39.0±0.6 |
| Q41203 | Jan. 4, 2012 | 3h43m43.0s | -4.0 | 8.85±0.91 | 107.1 | 75.5 | 232.2±0.1 | 49.7±0.1 | 42.5±0.3 | 40.7±0.3 | 39.2±0.3 |
| Q41204 | Jan. 4, 2012 | 4h03m14.7s | 2.0 | 0.020±0.003 | 95.3 | 84.2 | 229.7±0.1 | 48.5±0.1 | 42.9±0.5 | 41.2±0.5 | 38.5±0.5 |
| Q41205 | Jan. 4, 2012 | 4h04m05.5s | 0.0 | 0.157±0.019 | 99.1 | 91.9 | 234.2±0.1 | 47.7±0.1 | 42.7±0.6 | 40.9±0.6 | 39.1±0.6 |
| Q41206 | Jan. 4, 2012 | 4h31m05.0s | -0.5 | 0.229±0.003 | 103.4 | 96.5 | 229.9±0.2 | 48.1±0.1 | 43.9±0.5 | 42.2±0.5 | 38.8±0.5 |
| Q41207 | Jan. 4, 2012 | 4h43m09.8s | -0.5 | 0.271±0.031 | 103.7 | 85.6 | 229.9±0.2 | 48.1±0.1 | 44.0±0.5 | 42.3±0.5 | 39.2±0.5 |
| Q41208 | Jan. 4, 2012 | 4h54m53.3s | 1.0 | 0.062±0.007 | 99.9 | 88.6 | 230.9±0.2 | 49.9±0.1 | 42.3±0.5 | 40.6±0.5 | 38.8±0.5 |
| Q41209 | Jan. 4, 2012 | 4h56m36.1s | -6.0 | 52±6 | 104.9 | 71.1 | 232.4±0.2 | 49.7±0.1 | 42.0±0.5 | 40.2±0.5 | 38.8±0.5 |
| Q41210 | Jan. 4, 2012 | 5h01m21.5s | -1.0 | 0.482±0.053 | 101.7 | 83.9 | 228.9±0.1 | 48.7±0.1 | 43.7±0.5 | 40.0±0.5 | 39.0±0.5 |
| Q41211 | Jan. 4, 2012 | 5h06m46.5s | 1.0 | 0.049±0.006 | 98.5 | 92.6 | 228.8±0.1 | 50.1±0.1 | 42.9±0.6 | 41.2±0.6 | 38.9±0.6 |
| Q41212 | Jan. 4, 2012 | 5h08m00.2s | 0.0 | 0.142±0.021 | 99.8 | 89.9 | 228.9±0.2 | 48.8±0.1 | 43.6±0.7 | 41.9±0.7 | 38.9±0.7 |
| Q41213 | Jan. 4, 2012 | 5h08m40.7s | 2.0 | 0.026±0.004 | 103.7 | 87.3 | 228.5±0.1 | 49.3±0.1 | 43.2±0.4 | 41.5±0.4 | 38.7±0.4 |
| Q41214 | Jan. 4, 2012 | 5h08m52.2s | -1.5 | 0.721±0.082 | 98.9 | 82.1 | 230.2±0.1 | 49.2±0.1 | 42.6±0.7 | 40.9±0.7 | 38.2±0.7 |
| Q41215 | Jan. 4, 2012 | 5h08m57.3s | 3.0 | 0.008±0.001 | 98.9 | 85.0 | 229.3±0.5 | 49.1±0.2 | 43.2±0.6 | 41.5±0.6 | 38.8±0.6 |
| Q41216 | Jan. 4, 2012 | 5h09m34.9s | -0.5 | 0.273±0.029 | 99.1 | 85.2 | 231.6±0.3 | 48.4±0.2 | 42.3±0.5 | 40.6±0.5 | 38.4±0.5 |
| Q41217 | Jan. 4, 2012 | 5h12m36.6s | -4.0 | 8.6±0.9 | 98.7 | 80.1 | 233.3±0.1 | 49.1±0.1 | 42.2±0.5 | 40.5±0.5 | 39.0±0.5 |
| Q41218 | Jan. 4, 2012 | 5h15m20.1s | -3.0 | 2.5±0.3 | 95.2 | 79.9 | 230.3±0.1 | 47.5±0.1 | 43.5±0.6 | 41.8±0.6 | 38.7±0.6 |
| Q41219 | Jan. 4, 2012 | 5h16m59.3s | 0.0 | 0.141±0.021 | 100.9 | 92.3 | 228.4±0.3 | 47.5±0.2 | 44.0±0.6 | 42.3±0.6 | 38.6±0.6 |
| Q41220 | Jan. 4, 2012 | 5h18m10.1s | 0.5 | 0.098±0.012 | 100.7 | 90.6 | 229.6±0.2 | 49.0±0.1 | 42.9±0.6 | 41.2±0.6 | 38.6±0.6 |
| Q41221 | Jan. 4, 2012 | 5h21m59.0s | -5.0 | 21±2 | 109.8 | 79.7 | 232.4±0.3 | 49.5±0.1 | 42.2±0.6 | 40.5±0.6 | 38.9±0.6 |
| Q41222 | Jan. 4, 2012 | 5h36m42.1s | 1.0 | 0.049±0.006 | 97.7 | 88.6 | 230.3±0.3 | 48.4±0.1 | 43.1±0.6 | 41.4±0.6 | 38.7±0.6 |
| Q41223 | Jan. 4, 2012 | 5h36m48.2s | -3.0 | 3.0±0.4 | 98.2 | 81.0 | 228.9±0.2 | 50.1±0.1 | 42.7±0.6 | 41.0±0.6 | 38.8±0.6 |
| Q41224 | Jan. 4, 2012 | 5h41m11.1s | -0.5 | 0.244±0.031 | 105.7 | 91.1 | 229.6±0.2 | 48.9±0.1 | 43.9±0.6 | 41.2±0.6 | 38.6±0.6 |
| Q41225 | Jan. 4, 2012 | 5h52m50.4s | -2.0 | 0.942±0.120 | 98.5 | 77.0 | 229.6±0.2 | 48.9±0.1 | 43.6±0.6 | 41.9±0.6 | 38.7±0.6 |
| Q41226 | Jan. 4, 2012 | 5h59m49.0s | -1.0 | 0.405±0.051 | 103.3 | 80.6 | 233.6±0.9 | 48.2±0.4 | 41.9±0.7 | 41.2±0.7 | 38.6±0.7 |
| Q41227 | Jan. 4, 2012 | 6h07m27.8s | -4.5 | 13.0±1.7 | 98.3 | 75.4 | 229.9±0.2 | 48.2±0.1 | 43.0±0.5 | 41.3±0.5 | 38.5±0.5 |
| Q41228 | Jan. 4, 2012 | 6h10m08.0s | -0.5 | 0.256±0.031 | 102.4 | 94.1 | 232.7±0.9 | 49.3±0.4 | 42.3±0.6 | 40.6±0.6 | 39.0±0.6 |
| Q41229 | Jan. 4, 2012 | 6h11m50.6s | 1.0 | 0.052±0.006 | 98.1 | 89.3 | 230.3±0.2 | 48.7±0.1 | 43.6±0.6 | 40.9±0.6 | 38.5±0.6 |
| Q41230 | Jan. 4, 2012 | 6h12m00.6s | -1.0 | 0.310±0.040 | 103.7 | 86.9 | 228.8±0.1 | 48.0±0.1 | 43.5±0.6 | 41.8±0.6 | 38.6±0.6 |
| Q41231 | Jan. 4, 2012 | 6h13m03.4s | -4.0 | 7.25±0.82 | 102.5 | 71.2 | 228.6±0.2 | 48.4±0.1 | 43.6±0.5 | 42.0±0.5 | 38.1±0.5 |
| Q41232 | Jan. 4, 2012 | 6h18m57.8s | -3.0 | 2.17±0.30 | 97.0 | 80.6 | 227.0±0.2 | 48.5±0.2 | 43.9±0.6 | 42.6±0.6 | 38.7±0.6 |
| Q41233 | Jan. 4, 2012 | 6h19m47.7s | -1.0 | 0.322±0.038 | 96.8 | 80.8 | 228.6±0.1 | 48.1±0.1 | 43.8±0.5 | 42.2±0.5 | 38.8±0.5 |
| Q41234 | Jan. 4, 2012 | 6h25m14.5s | 0.0 | 0.148±0.020 | 97.2 | 85.6 | 230.0±0.2 | 48.3±0.1 | 42.9±0.6 | 41.2±0.6 | 38.5±0.6 |
| Q41235 | Jan. 4, 2012 | 6h31m22.2s | -3.0 | 2.51±0.29 | 98.5 | 77.6 | 227.4±0.03 | 49.9±0.02 | 43.1±0.4 | 41.5±0.4 | 38.7±0.4 |
| Q41236 | Jan. 4, 2012 | 6h32m20.7s | -0.5 | 0.217±0.029 | 99.8 | 86.6 | 229.78±0.03 | 50.53±0.02 | 42.5±0.6 | 40.8±0.6 | 39.0±0.6 |
| Q41237 | Jan. 4, 2012 | 6h35m00.0s | 0.0 | 0.138±0.019 | 101.2 | 86.0 | 231.8±0.1 | 46.9±0.1 | 43.2±0.5 | 41.5±0.5 | 38.7±0.5 |
| Q41238 | Jan. 4, 2012 | 6h39m52.5s | -5.0 | 23±3 | 100.1 | 75.2 | 233.7±0.1 | 52.0±0.1 | 41.0±0.6 | 39.3±0.6 | 39.2±0.6 |
| Q41239 | Jan. 4, 2012 | 6h44m30.6s | -1.5 | 0.579±0.061 | 99.6 | 89.1 | 230.3±0.1 | 47.3±0.1 | 43.6±0.5 | 42.0±0.5 | 38.7±0.5 |

Table 3. Orbital elements (J2000) for 85 multi-station Quadrantids recorded from 2008 to 2012.

| Meteor Code | a (AU) | e | i (°) | Ω (°) ±10$^{-4}$ | ω (°) | q (AU) |
|---|---|---|---|---|---|---|
| Q40800 | 2.9±0.3 | 0.67±0.03 | 73.8±0.5 | 282.9859 | 171.0±0.3 | 0.9784±0.0003 |
| Q40801 | 3.2±0.4 | 0.69±0.04 | 71.7±0.7 | 282.9872 | 169.1±0.5 | 0.9760±0.0005 |
| Q40802 | 2.7±0.3 | 0.65±0.03 | 72.5±0.5 | 282.9899 | 156.9±0.7 | 0.952±0.001 |
| Q40803 | 2.9±0.1 | 0.66±0.01 | 71.6±0.3 | 282.9932 | 167.2±0.2 | 0.9736 ±0.0002 |
| Q40804 | 3.1±0.3 | 0.69±0.03 | 71.0±0.5 | 283.0023 | 165.6±0.5 | 0.9707±0.0006 |
| Q40805 | 3.0±0.3 | 0.67±0.03 | 73.7±0.6 | 283.0045 | 174.3±0.7 | 0.9813±0.0004 |
| Q40806 | 3.0±0.2 | 0.67±0.03 | 72.4±0.5 | 283.0123 | 169.3±0.4 | 0.9764±0.0003 |
| Q40807 | 2.7±0.3 | 0.63±0.03 | 74.7±0.7 | 283.0204 | 172.1±0.7 | 0.9796±0.0006 |
| Q40808 | 2.8±0.2 | 0.64±0.03 | 71.4±0.5 | 283.0230 | 171.2±0.4 | 0.9787±0.0003 |
| Q40809 | 3.2±0.3 | 0.69±0.03 | 72.8±0.5 | 283.0247 | 174.3 ±0.3 | 0.9813 ±0.0002 |
| Q40810 | 2.9±0.2 | 0.66±0.03 | 74.2±0.5 | 283.0304 | 171.1±0.4 | 0.9786±0.0004 |
| Q40811 | 2.8±0.2 | 0.65±0.03 | 70.6±0.5 | 283.0354 | 169.4±0.4 | 0.9766±0.0003 |
| Q40812 | 2.7±0.3 | 0.64±0.03 | 72.1±0.6 | 283.0387 | 168.9±0.5 | 0.9760±0.0005 |
| Q40813 | 3.0±0.2 | 0.67±0.02 | 72.2±0.4 | 283.0407 | 171.0±0.4 | 0.9784±0.0003 |
| Q40814 | 3.0±0.2 | 0.67±0.02 | 73.1±0.4 | 283.0415 | 169.5±0.4 | 0.9767±0.0004 |
| Q40815 | 2.9±0.3 | 0.67±0.03 | 70.3±0.5 | 283.0426 | 163.9±0.5 | 0.9679±0.0006 |
| Q40816 | 2.9±0.2 | 0.66±0.03 | 71.3±0.5 | 283.0476 | 168.9±0.4 | 0.9760±0.0003 |
| Q40817 | 2.9±0.3 | 0.67±0.04 | 71.1±0.6 | 283.0477 | 173.4±0.4 | 0.9807±0.0003 |
| Q40818 | 2.8±0.3 | 0.65±0.03 | 73.1±0.6 | 283.0495 | 165.9±0.5 | 0.9716 ±0.0006 |
| Q40819 | 3.1±0.3 | 0.68±0.03 | 72.9±0.5 | 283.0528 | 166.0±0.5 | 0.9714±0.0006 |
| Q40820 | 3.3±0.4 | 0.70±0.03 | 71.8±0.6 | 283.0547 | 170.2±0.4 | 0.9773±0.0004 |
| Q21000 | 2.5±0.3 | 0.62±0.04 | 75.6±0.8 | 281.5552 | 178.8±0.9 | 0.9832±0.0001 |
| Q31100 | 2.8±0.3 | 0.65±0.03 | 73.5±0.5 | 282.8850 | 171.0±0.4 | 0.9785±0.0003 |
| Q31101 | 2.9±0.2 | 0.66±0.03 | 70.3±0.5 | 283.0422 | 165.8±0.5 | 0.9714±0.0005 |
| Q31102 | 4.4±0.6 | 0.77±0.03 | 71.6±0.5 | 283.0957 | 173.5±1.0 | 0.9805±0.0008 |
| Q31103 | 2.4±0.6 | 0.59±0.03 | 71.7±0.6 | 283.0895 | 173.5±1.0 | 0.9814±0.0007 |
| Q41100 | 3.0±0.3 | 0.68±0.03 | 71.8±0.5 | 283.1165 | 168.3±0.4 | 0.9751±0.0004 |
| Q41101 | 2.1±0.1 | 0.54±0.03 | 77.4±0.6 | 283.1221 | 184.3±0.8 | 0.9823±0.0004 |
| Q41102 | 3.1±0.4 | 0.69±0.04 | 68.8±0.6 | 283.1288 | 171.1±0.6 | 0.9784±0.0006 |
| Q41103 | 2.8±0.2 | 0.65±0.02 | 69.4±0.4 | 283.1323 | 174.5±0.5 | 0.9815±0.0003 |
| Q41104 | 2.6±0.1 | 0.62±0.02 | 69.8±0.3 | 283.1330 | 172.6±0.6 | 0.9801±0.0004 |
| Q41105 | 3.4±0.5 | 0.71±0.03 | 71.8±0.6 | 283.1470 | 178.5±1.0 | 0.9831±0.0002 |
| Q41106 | 3.2±0.4 | 0.69±0.03 | 71.8±0.6 | 283.1552 | 176.5±0.9 | 0.9825±0.0004 |
| Q41107 | 2.6±0.1 | 0.62±0.02 | 71.9±0.3 | 283.1399 | 171.7±0.6 | 0.9793±0.0005 |
| Q41108 | 3.0±0.2 | 0.67±0.02 | 73.7±0.4 | 283.1863 | 169.1±0.5 | 0.9761±0.0006 |
| Q41109 | 3.6±0.2 | 0.73±0.03 | 70.0±0.3 | 283.1446 | 175.1±0.5 | 0.9823±0.0002 |
| Q41110 | 2.9±0.2 | 0.66±0.03 | 72.9±0.5 | 283.1935 | 174.0±0.5 | 0.9812±0.0004 |
| Q41111 | 2.9±0.3 | 0.66±0.3 | 71.3±0.6 | 283.1966 | 177.1±0.6 | 0.9828±0.0002 |
| Q41112 | 3.1±0.3 | 0.69±0.03 | 70.1±0.5 | 283.2210 | 163.2±0.4 | 0.9662±0.0005 |
| Q41113 | 3.3±0.2 | 0.70±0.02 | 71.1±0.3 | 283.2458 | 174.9±0.5 | 0.9816±0.0003 |
| Q41114 | 3.4±0.2 | 0.71±0.02 | 69.4±0.3 | 283.2485 | 176.6±0.4 | 0.9825±0.0002 |
| Q41115 | 3.1±0.3 | 0.68±0.03 | 72.0±0.6 | 283.2494 | 167.4±0.5 | 0.9737±0.0005 |
| Q41116 | 3.4±0.5 | 0.71±0.04 | 68.6±0.7 | 283.2992 | 167.6±0.5 | 0.9737±0.0004 |
| Q41117 | 3.4±0.2 | 0.71±0.02 | 69.4±0.3 | 283.3559 | 176.5±0.5 | 0.9825±0.0002 |
| Q41118 | 2.4±0.2 | 0.59±0.03 | 69.8±0.6 | 283.3717 | 172.2±1.2 | 0.9799±0.0010 |
| Q41200 | 3.2±0.4 | 0.70±0.04 | 72.9±0.7 | 282.9432 | 169.8±0.5 | 0.9770±0.0005 |
| Q41201 | 3.1±0.3 | 0.68±0.03 | 72.3±0.6 | 282.9547 | 166.9±0.5 | 0.9730±0.0005 |
| Q41202 | 3.2±0.4 | 0.69±0.04 | 69.5±0.7 | 283.0076 | 168.6±0.6 | 0.9754±0.0007 |
| Q41203 | 3.3±0.2 | 0.70±0.01 | 70.4±0.3 | 283.0099 | 169.9±0.3 | 0.9770±0.0003 |
| Q41204 | 2.7±0.3 | 0.64±0.03 | 72.3±0.6 | 283.0237 | 169.9±0.5 | 0.9773±0.0004 |
| Q41205 | 3.2±0.3 | 0.70±0.03 | 70.8±0.5 | 283.0243 | 163.5±0.5 | 0.9666±0.0009 |
| Q41206 | 2.9±0.3 | 0.67±0.03 | 74.1±0.5 | 283.0434 | 166.1±0.5 | 0.9717±0.0005 |
| Q41207 | 3.3±0.3 | 0.70±0.03 | 73.7±0.5 | 283.0519 | 169.4±0.4 | 0.9764±0.0004 |
| Q41208 | 2.9±0.3 | 0.67±0.03 | 70.6±0.5 | 283.0603 | 171.2±0.4 | 0.9787±0.0003 |
| Q41209 | 3.0±0.3 | 0.67±0.03 | 69.8±0.5 | 283.0615 | 169.2±0.3 | 0.9763±0.0003 |
| Q41210 | 3.1±0.3 | 0.68±0.03 | 73.4±0.5 | 283.0648 | 171.7±0.4 | 0.9790±0.0003 |
| Q41211 | 3.0±0.3 | 0.67±0.03 | 71.8±0.6 | 283.0687 | 174.1±0.3 | 0.9812±0.0002 |
| Q41212 | 3.1±0.4 | 0.68±0.04 | 73.2±0.7 | 283.0695 | 171.8±0.4 | 0.9792±0.0003 |
| Q41213 | 2.9±0.2 | 0.66±0.02 | 72.7±0.5 | 283.0700 | 173.0±0.3 | 0.9804±0.0002 |
| Q41214 | 2.8±0.4 | 0.66±0.04 | 71.5±0.7 | 283.0702 | 170.6±0.4 | 0.9781±0.0003 |
| Q41215 | 3.0±0.3 | 0.67±0.03 | 72.5±0.7 | 283.0702 | 171.7±0.7 | 0.9791±0.0007 |
| Q41216 | 2.7±0.2 | 0.64±0.03 | 71.1±0.6 | 283.0707 | 167.3±0.5 | 0.9739±0.0006 |
| Q41217 | 3.1±0.3 | 0.69±0.03 | 70.0±0.5 | 283.0729 | 167.1±0.4 | 0.9732±0.0004 |
| Q41218 | 2.9±0.3 | 0.66±0.03 | 73.3±0.6 | 283.0747 | 167.4±0.5 | 0.9739±0.0004 |
| Q41219 | 2.8±0.3 | 0.65±0.03 | 74.6±0.7 | 283.0758 | 169.8±0.7 | 0.9771±0.0007 |
| Q41220 | 2.8±0.3 | 0.66±0.03 | 72.2±0.6 | 283.0767 | 107.0±0.4 | 0.9785±0.0003 |
| Q41221 | 3.1±0.3 | 0.68±0.03 | 70.2±0.7 | 283.0795 | 168.9±0.5 | 0.9758±0.0006 |
| Q41222 | 2.9±0.3 | 0.66±0.03 | 72.5±0.6 | 283.0898 | 169.1±0.4 | 0.9763±0.0004 |
| Q41223 | 2.9±0.3 | 0.67±0.03 | 71.6±0.6 | 283.0899 | 174.1±0.3 | 0.9812±0.0002 |
| Q41224 | 2.8±0.3 | 0.65±0.03 | 72.3±0.6 | 283.0930 | 170.8±0.4 | 0.9782±0.0003 |
| Q41225 | 2.9±0.3 | 0.66±0.03 | 73.6±0.6 | 283.1012 | 173.1±0.3 | 0.9805±0.0002 |
| Q41226 | 2.8±0.4 | 0.65±0.04 | 70.0±0.9 | 283.1063 | 164.6±1.4 | 0.969±0.002 |
| Q41227 | 2.7±0.2 | 0.64±0.03 | 72.7±0.5 | 283.1116 | 169.2±0.4 | 0.9765±0.0004 |
| Q41228 | 3.2±0.4 | 0.69±0.04 | 70.3±0.7 | 283.1136 | 168.2±1.2 | 0.974±0.002 |
| Q41229 | 2.7±0.3 | 0.64±0.03 | 71.8±0.6 | 283.1147 | 169.7±0.5 | 0.9770±0.0004 |
| Q41230 | 2.8±0.3 | 0.65±0.03 | 73.7±0.6 | 283.1148 | 170.3±0.4 | 0.9777±0.0003 |
| Q41231 | 2.9±0.3 | 0.66±0.03 | 73.7±0.5 | 283.1155 | 171.2±0.3 | 0.9787±0.0002 |
| Q41232 | 2.8±0.3 | 0.66±0.04 | 74.5±0.6 | 283.1197 | 173.7±0.4 | 0.9809±0.0002 |
| Q41233 | 3.0±0.3 | 0.67±0.03 | 74.0±0.5 | 283.1203 | 170.9±0.3 | 0.9783±0.0003 |
| Q41234 | 2.8±0.3 | 0.64±0.03 | 72.5±0.6 | 283.1242 | 169.2±0.4 | 0.9765±0.0004 |
| Q41235 | 2.9±0.2 | 0.66±0.02 | 72.7±0.4 | 283.1285 | 175.5±0.1 | 0.9821±0.0001 |
| Q41236 | 3.1±0.4 | 0.68±0.03 | 70.9±0.6 | 283.1293 | 173.8±0.2 | 0.9810±0.0001 |
| Q41237 | 2.9±0.3 | 0.66±0.03 | 72.8±0.5 | 283.1311 | 164.3±0.5 | 0.9687±0.0005 |
| Q41238 | 3.4±0.4 | 0.71±0.03 | 67.3±0.6 | 283.1347 | 172.3±0.3 | 0.9796±0.0002 |
| Q41239 | 2.9±0.3 | 0.67±0.03 | 73.7±0.5 | 283.1378 | 167.0±0.4 | 0.9732±0.0004 |

Table 4. Aerodynamic pressure for flares and break-up processes discussed in the text.

| Meteor code | Height (km) | Velocity (km $s^{-1}$) | Aerodynamic pressure (dyn $cm^{-2}$) |
|---|---|---|---|
| Q21000 | 93 ± 1 | 43.1 ± 0.8 | $(2.7 \pm 0.7)\cdot 10^4$ |
| Q41110 | 79 ± 1 | 43.3 ± 0.5 | $(3.5 \pm 0.3)\cdot 10^5$ |
| Q41118 | 82 ± 1 | 39.5 ± 0.5 | $(1.7 \pm 0.3)\cdot 10^5$ |